# Perovskite nanocrystal self-assemblies in 3D hollow templates

*Etsuki Kobiyama[1], Darius Urbonas[1], Maryna I. Bodnarchuk[2,3], Gabriele Rainò[2,3], Antonis Olziersky[1], Daniele Caimi[1], Marilyne Sousa[1], Rainer F. Mahrt[1], Maksym V. Kovalenko[2,3], Thilo Stöferle[1].*

[1] IBM Research Europe – Zurich, Rüschlikon, Switzerland

[2] Institute of Inorganic Chemistry, Department of Chemistry and Applied Bioscience, ETH Zurich, Zurich, Switzerland

[3] Laboratory of Thin Films and Photovoltaics, Empa — Swiss Federal Laboratories for Materials Science and Technology, Dübendorf, Switzerland

Corresponding Authors: maryna.bodnarchuk@empa.ch, mvkovalenko@ethz.ch, tof@zurich.ibm.com

**Abstract**

Highly ordered nanocrystal (NC) assemblies, namely superlattices (SLs), have been investigated as novel building blocks of optical and optoelectronic devices due to their unique properties based on interactions among neighboring NCs. In particular, lead halide perovskite NC SLs have attracted significant attention, owing to their extraordinary optical characteristics of individual NCs and collective emission processes like superfluorescence (SF). So far, the primary method for preparing perovskite NC SLs has been the drying-mediated self-assembly method, in which the colloidal NCs spontaneously assemble into SLs during solvent evaporation. However, this method lacks controllability because NCs form random-sized SLs at random positions on the substrate rendering NC assemblies in conjunction with device structures such as photonic waveguides or microcavities challenging. Here, we demonstrate template-assisted self-assembly to deterministically place perovskite NC SLs and control their geometrical properties . A solution of $CsPbBr_3$ NCs is drop-casted on a substrate with lithographically-defined hollow structures. After solvent evaporation and removal of excess NCs from the substrate surface, NCs only remain in the templates thereby defining the position and size of these NC assemblies. We performed photoluminescence (PL) measurements on these NC assemblies and observed signatures of SF, similar as in spontaneously assembled SLs. Our findings are crucial for optical devices that harness embedded perovskite NC assemblies and prepare fundamental studies on how these collective effects can be tailored through the SL geometry.

Colloidal semiconductor nanocrystals are often referred to as "artificial atoms" or "quantum dots" since their physical properties originate from quantized electronic states.[1,2] Because of the outstanding optical properties and broad range of potential applications, extensive research has been carried out. Thanks to the recent advances in synthesis technology, nanocrystal (NC) ensembles with very narrow size distribution can be produced in easy and cost-effective methods,[3] allowing to tailor meso-structured materials by using individual NCs as building blocks. One of the best known meso-structured materials made out of colloidal NCs are highly ordered NC solids, so called NC superlattices (SLs).[4,5] SLs exhibit not only physical properties arising from individual NCs but also unique properties based on inter-NC interactions in analogy to characteristic properties of bulk solids which arise from inter atomic interactions.[6–8] Because it is possible to engineer the optical and electronic properties of SLs by altering their composition, NC SLs have been investigated to create materials with tailored physical properties.

One of the most striking impacts of SLs on the optical properties is the emergence of cooperative photon emission resulting from enhanced inter-NC interactions. Due to the closely packed arrangement of NCs, an excited electric dipole in a NC can coherently interact with dipoles excited in neighboring NCs. One typical cooperative photon emission process, namely superfluorescence (SF),[9,10] has been observed from SLs consisting of lead halide perovskite NCs.[11–13] Due to its characteristic fast and intense emission with narrow linewidth, SF has been of interest for ultrafast photonic technology as well as for novel bright (quantum) light sources.

Thus far, the most used method for SL fabrication is the drying-mediated self-assembly method.[5,14] In this process, NCs spontaneously arrange into highly ordered structures while the solvent slowly evaporates. However, because nucleation occurs at random once when the saturation density is reached locally during solvent evaporation, the position and size of SLs cannot be controlled. Therefore, it has been challenging to deterministically position and shape SLs in order to integrate them into micro- or nano-device structures such as photonic waveguides or resonators.

For unordered colloidal NC assemblies, different template-assisted deposition techniques have been developed in the recent decades. These ranged from deposition of individual NC in arrays[15,16] through convective, capillary assembly over deterministic, number- and geometry-controlled assemblies[17–19] to large clusters in photonically functional nanostructures.[20–22] Nevertheless, a controlled template-assisted assembly of SLs exhibiting cooperative optical emission is missing.

Here, we introduce hollow templates with three-dimensional confinement and demonstrate template-assisted self-assembly to achieve precise positioning and size definition of perovskite NC SLs. We study the effect of different solvents, NC ligand molecules, and template geometries on the assembling process and yield. Additionally, using time-resolved spectroscopy at cryogenic temperatures, we optically characterize the SLs and observe signatures of SF that are consistent with results from spontaneously assembled ordered SLs[11,12] but clearly distinct from spin-coated NC films.

**Results and Discussion**

For the template structure fabrication, we adapted the processes developed for template-assisted selective epitaxial growth of III-V semiconductors on silicon.[23] The fabrication steps of the template structures are sketched in **Figure 1**a. (i) Silicon-on-insulator (SOI) substrates or glass substrates with a Si layer on top were patterned using e-beam lithography (EBL) and reactive ion etching (RIE). The Si layer has a thickness of 220 nm defining the height of the hollow space within the template structures. (ii) To realize the walls and ceiling of the templates, conformal atomic layer deposition (ALD) and electron-beam evaporation were used to deposit a 200 nm thick

SiO$_2$ layer over the structures. (iii) Template openings were defined by EBL and reactive ion etching of the encapsulating SiO$_2$ layer. (iv) The exposed Si layer was dry-etched with XeF$_2$ gas. A scanning electron microscopy (SEM) image of the final template structure is shown in **Figure 1**b. (v) After the templates were constructed, a CsPbBr$_3$ NC solution in toluene was drop-casted on the substrates, and the solvent slowly evaporated in the toluene atmosphere. CsPbBr$_3$ NCs with a mono-disperse size distribution (8.7 ± 0.6 nm) were produced leveraging the recently reported room-temperature synthesis platform based on PbBr$_2$/trioctylphosphine oxide (TOPO) molecular adducts as the precursor, wherein the formation of NCs is precisely adjustable owing to slower reaction kinetics.[24] The ligands used in the TOPO/PbBr$_2$ method are loosely bound TOPO and dialkyl phosphinate anion and, hence, are readily displaced with a ligand of choice. In this study, didodecyldimethylammonium bromide (DDAB) or oleic acid and oleylamine (OAc/OAm) were added at room temperature at the end of the NC formation. (see Methods and Supplementray Figure S1). (vi) Finally, excess NCs on the substrate surface are removed by applying an optics cleaning polymer (Red First Contact Polymer, Photonic Cleaning Technologies).

Inspection of the SL formation has been firstly conducted by an optical microscope. Bright field optical images show the hollow template structures in white color before the NC deposition, while the substrate is shown as light blue color (**Figure 1**c Top). After the NC drop-casting and the excess NCs removal processes, the templates are displayed in green color because they are partially or fully filled with green perovskite NCs (**Figure 1**c Bottom). Furthermore, we used a scanning transmission electron microscope (STEM) to examine the cross section of the template-assisted NC assemblies (see Methods). Within the templates, the NCs form ordered structures, as shown in **Figures 1**d and Supporting Figure S2. The NC layer is 60-70 nm thick, indicating an air gap between perovskite NC layer and the upper window and an incomplete filling of the templates by the nanocrystals It is worth noting that, because the sidewalls of the templates were removed and the top window dropped down after the STEM lamella preparation, the air gap is not visible in the STEM image.

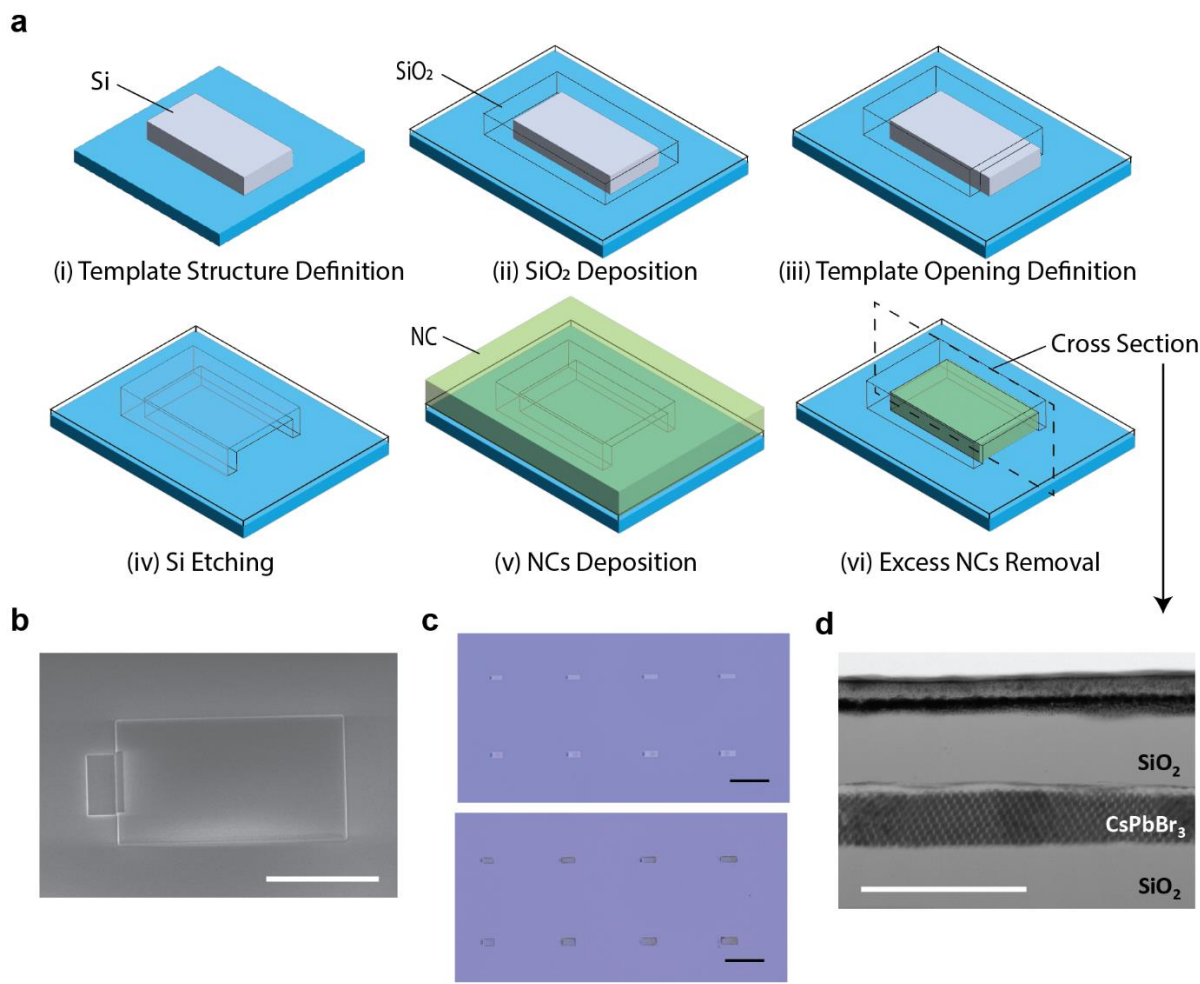

**Figure 1.** (a) Schematic illustration of the process flow to fabricate transparent, hollow templates that are subsequently filled with NCs. (b) SEM image of a template structure. Scale bar: 5 μm. (c) Optical images of an array of template structures before NCs deposition (Top), and template-assisted NC assemblies after NCs deposition (Bottom). Scale bar: 25 μm. (d) Bright field STEM image showing a cross section of a template assisted NC assembly. Scale bar: 200 nm.

**Influence of solvent and ligands.** Since the self-assembly process is driven by interactions at the interfaces of vapor-liquid, ligand-ligand, ligand-solvent, solvent-substrate/template and ligand-substrate/template, the surface characteristics of NCs, solvent, substrates and templates play an important role. However, the complexity of the liquid with its manifold of parameters (many of which are not known precisely) and its interplay with nanoscale confinement geometry renders accurate modelling difficult.[25] Therefore, we investigate empirically the influence of the main factors. First, we studied combinations of NC surface capping ligands and the solvent. **Figure 2**a-c show the absorption and PL spectra of different NC solutions. The solution in **Figure 2**a is comprised of a mixture of oleic acid (OA), oleylamine (OLA), and didodecyldimethylammonium bromide (DDAB) as capping ligands and toluene as solvent (denoted as solution A). The solution in **Figure 2**b consists of phosphatidylserine (Ptd-L-Ser) as capping ligands and toluene as solvent (denoted as solution B). The solution C in **Figure 2**c is comprised of DDAB as capping ligands and cycloheptane as solvent. These three different types of solutions exhibit identical absorption and PL spectra, which assures that the size distributions of the three different NC ensembles are the same. Besides, we can confirm that the absorption and PL spectra of the NCs are not affected by the ligand/ solvent exchange processes.

**Figure 2**d-f are optical images of typical template-assisted NC assemblies fabricated from the above mentioned three different solutions. From solution A, the NCs assembled homogeneously in the templates (**Figure 2**d), while NCs assembled into multiple (sub-)micron-sized clusters in the case of solution B (**Figure 2**e). The spatial (in)homogeneity in both cases is clearly visible in the spatially resolved PL (**Figure 2**g, h) map as well. The NC assembly from solution A shows homogeneous PL intensity over the assembly (**Figure 2**g). On the contrary, the NC assembly from solution B shows a very inhomogeneous image with several bright spots that correspond to small

clusters (**Figure 2**h), which appear a bit merged compared to the room temperature microscope image due to the lower resolution of the micro PL setup. Since the capping ligands (A: DDAB with OA and OLA, B: Ptd-L-Ser) are the only difference between solutions A and B, it is evident that the capping ligands play a significant role in the template-assisted self-assembly process. DDAB is preferred for producing homogenous NC assemblies, whereas the polydispersity of Ptd-L-Ser molecules is presumably the reason why NCs with Ptd-L-Ser ligands did not form homogeneous NC assemblies.

**Figure 2**f, i display the optical image and the results of the PL mapping measurements of a template device after the assembling process with solution C, respectively. The optical image does not show any greenish part, and the PL map does not show any PL signals. These results indicate that NCs are not present in the template structure. We attribute the lack of NCs to cycloheptane's low wettability to the hydrophilic $SiO_2$ surfaces due to its lower polarity than toluene, given the solvent difference between solutions A and C. Therefore, it is likely that the solution did not enter the hollow, sub-micron-high template structures. Considering the above observations, we concluded that DDAB and toluene are the preferable combination of capping ligands and solvent for the self-assembly in templates made from $SiO_2$.

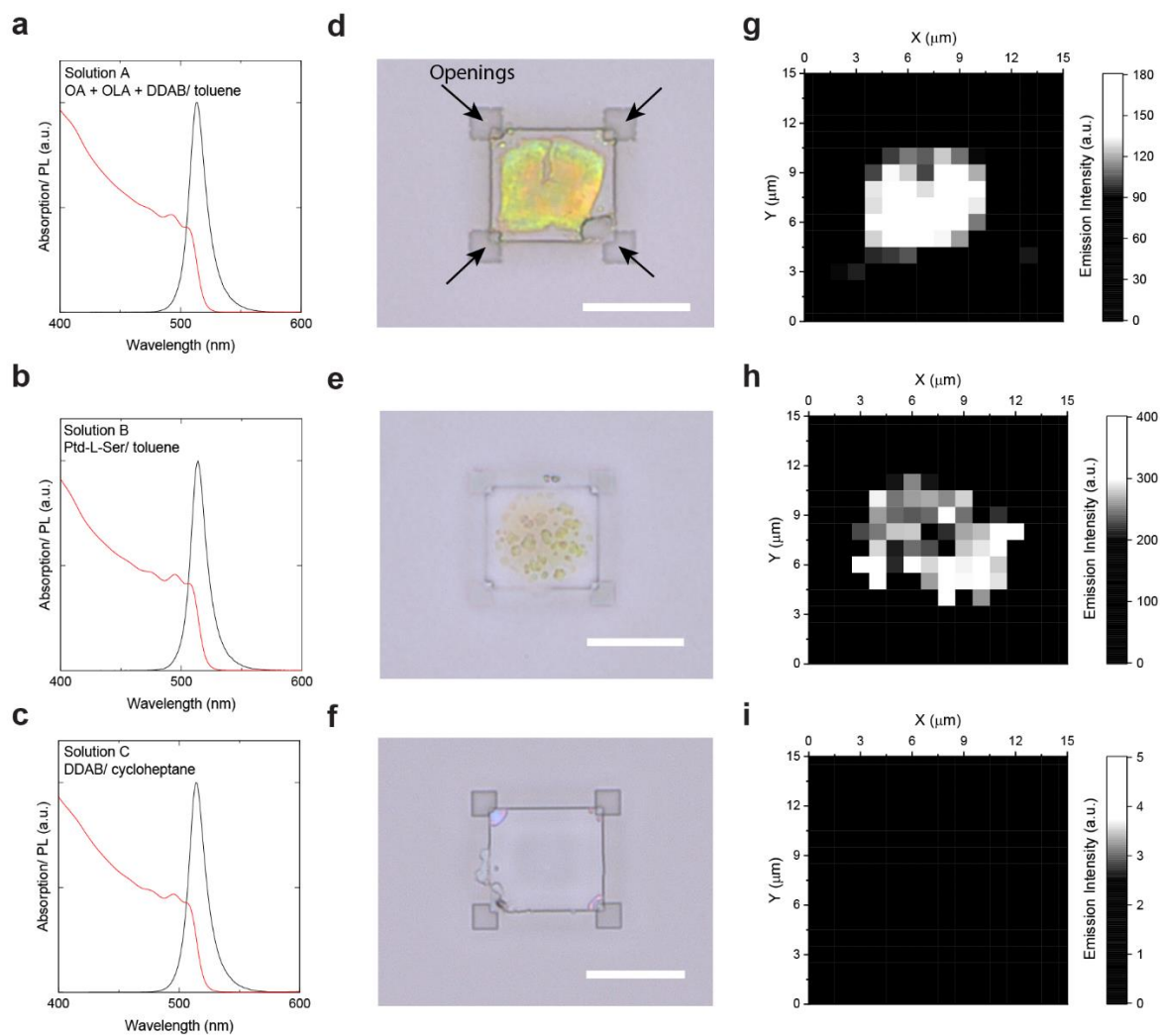

**Figure 2.** Absorption (red solid lines) and PL (black solid lines) spectra of NC solutions (a-c), optical microscope images (d-f, scale bars: 10 μm), and spatially resolved PL maps (g-i) of NC assemblies prepared by the template-assisted method. The surface capping ligands and solvent of the solution are: (a, d, g) ligands: oleic acid (OA) + oleylamine (OLA) + didodecyldimethylammonium bromide (DDAB), solvent: toluene, (b, e, h) ligands: phosphatidylserine (Ptd-L-Ser), solvent: toluene, (c, f, i) ligands: DDAB, solvent: cycloheptane.

**Optimizing the template geometry.** Here, we studied the effects of the template geometry on the assembling process of solution A. We evaluated the yield of the process by optical image analysis (see Methods). The masked images (Supporting Figure S3) display a white color to indicate the area filled with NCs, which was identified by analyzing the RGB values of individual pixels. From the image analysis, we obtained histograms of the filling ratio (**Figure 3**). To determine the ideal template design, the number of opening slits through which the NC solution is inserted into the template structures was varied, as each plot's inset illustrates. Each histogram shows the results of 100 template structures having rectangular shapes ranging in size from 1 to 100 $\mu m^2$ and aspect ratios from 0.1 to 10 (Supporting Figure S4a). The filled ratio is defined as:

$$\text{Filled ratio} = \frac{\text{Area filled with NCs}}{\text{Area of template structure (defined by EBL)}}$$

When the filled ratio equals 1, the template structure is perfectly filled by NCs, while for a filled ratio smaller than 1, the template structure is partially filled.

In the case of templates with one opening (**Figure 3**a), the mean value of the filled ratio is 0.54. On the other hand, the template structures with two openings (**Figure 3**b) and four openings (**Figure 3**c) result in the average filled ratio of 0.24 and 0.20, respectively. Besides, half of the structures have a filled ratio between 0.36 and 0.72 for one opening structure, while the corresponding values are 0.15 ~ 0.32 and 0.06 ~ 0.28 for two openings and four openings, respectively (Supporting Figure S4b). Based on the statistical analysis above, we conclude that one-opening template structures are the preferable design. Furthermore, we discover that elongated structures with a high aspect ratio are advantageous within the one-opening templates. (Supporting Figure S3f). These findings suggest that, in contrast to the other designs, a well-defined evaporation and deposition front can propagate during the assembly process through the elongated one-opening templates. It is noteworthy that the reduced physical robustness of the structures due to the smaller sidewall areas compared to one-opening templates, which occasionally cannot withstand the cleaning process with optics cleaning polymer, is another factor contributing to the low yields for two- or four-opening templates (step (vi) in **Figure 1**a).

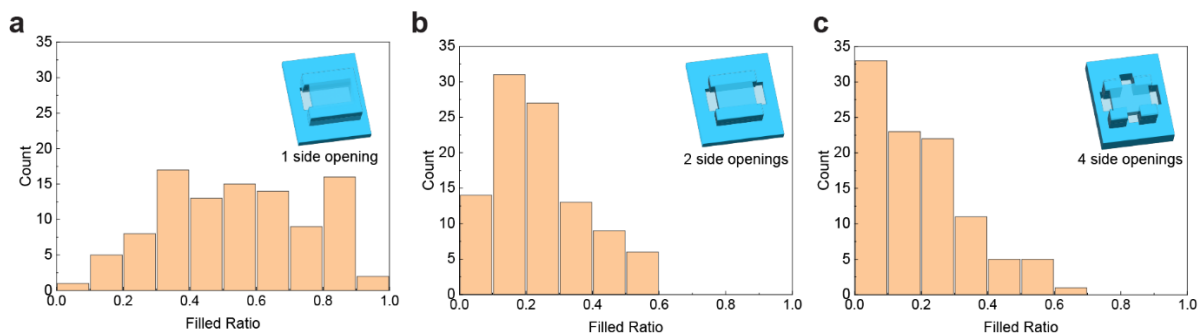

**Figure 3.** Statistical assembly yield determined from optical microscopic image analysis for different numbers of openings in the hollow template. (a-c) Histograms of the ratio of the area filled with NCs to the total area of the template structure for different numbers of openings: (a) on one side, (b) on two sides, and (c) on four sides. Insets show schematics of each template design.

**Superfluorescent emission.** Next, we investigated the optical properties of the obtained assemblies. We performed the optical characterization of template-assisted NC assemblies at 6 K because, at room temperature, the emission features are typically spectrally too broad to accurately evaluate the spatial homogeneity or spectral shifts from aggregates, and because cooperative, superfluorescent emission only happens at lower temperatures (see Methods). Firstly, we measured emission from different positions within a single template structure to check the homogeneity of the NC assembly (**Figure 4**a). We chose a typically filled template with one opening and dimensions of 10 μm × 10 μm, with the excitation spot size being 4.4 μm in diameter (FWe$^{-2}$M). As shown in **Figure 4**b, the spectral features are similar at all measured positions, consisting of two-peak structures with peak center wavelengths of 522~523 nm and 537~538 nm. The two-peak structure is consistent with the previously reported emission spectra of superfluorescent NC assemblies, where the high energy peak is assigned to the PL signal from an ensemble of uncoupled NCs, while the low energy peak is assigned to the coupled NC state.[11,26] The pronounced red tail between 540 nm and 560 nm might be due to defect states occurring after the assembly. Additionally, we noticed that there are no appreciable variations in emission dynamics across the measured positions based on time-resolved photoluminescence measurements (**Figure 4**c). In the time range of 0 ~ 1.0 ns, the PL time traces exhibit a single exponential decay with a lifetime of 300 ± 50 ps, consistent with the low-temperature PL decay of these CsPbBr$_3$ NCs in the low-intensity, non-SF regime. The longer tail may be related to trapping or defect states-mediated delay fluorescence.[27,28]

For increasingly stronger excitation pulses, we observed a significant change in emission spectra and dynamics. **Figure 5**a shows the emission spectra from a single measurement position of a template-assisted NC assembly under increasing excitation fluences. The emission peak at 537 nm from the coupled NCs shows a drastic intensity increase with increasing excitation fluence. As shown in **Figure 5**b, the time-integrated emission intensity of the red-shifted emission peak (wavelength range 535 ~ 570 nm) shows a linear dependence on excitation fluence, reflecting the fact that there are no excitation density-dependent non-radiative processes in the system. From the time-resolved measurements, we observed a continuous shortening of emission lifetime (**Figure 5**c and **5**d top panel). The peak intensity shows a super-linear dependence on excitation fluence (**Figure 5**d bottom panel). These spectral and temporal dynamics, which are dependent on excitation fluence, are consistent with SF emission, as observed in spontaneously assembled ordered SLs without templates.[11]

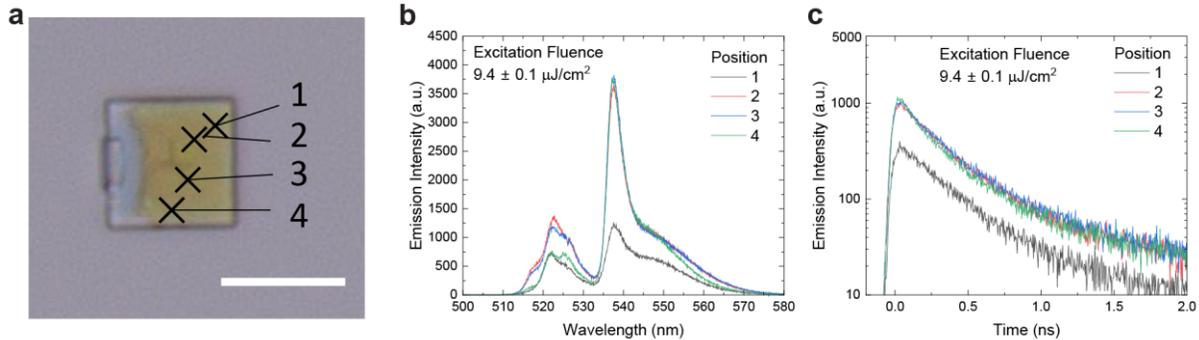

**Figure 4.** Spatially resolved PL from a NC assembly. (a) Optical microscopic image of the measured template assembly. Crosses indicate the measured positions. (Scale bar: 10 μm) (b) PL spectra and (c) time traces obtained at 6 K from the four different positions in the assembly.

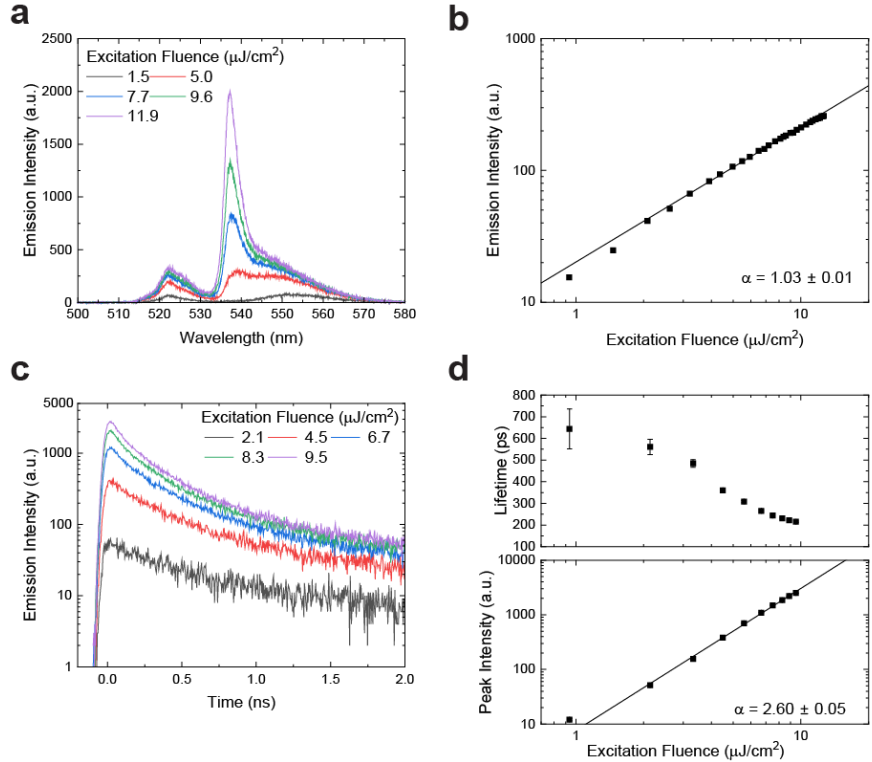

**Figure 5.** Excitation fluence dependence of PL from a NC assembly at 6 K. (a) PL spectra and (b) integrated emission intensity as a function of excitation fluence. α indicates the exponent value obtained from a power-law fit (solid line). The integrated wavelength range is from 535 to 570 nm. (c) PL time traces for different excitation fluences. The PL signal within the wavelength range from 532 to 548 nm was selected with a band-pass filter. (d) Emission lifetime (top panel) and emission peak intensity (bottom panel) as functions of excitation fluence. α indicates the exponent value obtained from a power-law fit (solid line).

To scrutinize the ultrafast emission dynamics of SF at even higher excitation fluence, we performed time-resolved measurements using an amplified femtosecond laser system and a streak camera instead of a pulsed diode laser with an avalanche photodiode (see Methods). Here, the excitation beam covers the whole template precluding spatially-resolved analysis. As shown in **Figure 6**a, the observed red-shifted emission peak with respect to the individual NCs emission energy was similar as for the microscopic optical characterization (**Figure 4**b and **Figure 5**a). In contrast, the emission spectra of a spincoated NC film (**Figure 6**b) and a NC assembly by drying mediated method (Supporting Figure S5a) mainly consist of two emission peaks with energy separation of 23 ~ 26 meV, comparable with the reported values of energy separation of excitons and trions.[29,30] In addition, for the higher excitation fluences, we observed narrow emission peaks between 530 and 540 nm in the emission spectra from the drying mediated NC assembly. We attributed these regularly spaced emission peaks to whispering gallery lasing modes from the internal reflection within the NC assembly. We speculate that the difference between the spectral shapes of template-assisted NC assemblies and drying mediated NC assemblies is caused by the cleaning process (**Figure 1**a (vi)). As shown in Supporting Figure S5b, drying mediated NC assemblies also show the pronounced red-shifted emission peak at 540 nm after the cleaning polymer solution is applied.

In terms of the temporal emission dynamics, we observed emission pulse ringing in the time domain (**Figure 6**c), so called Burnham Chiao ringing,[31] which is a fingerprint of the SF regime. On the other hand, such a pronounced pulse ringing effect was not observed in spincoated NC films (**Figure 6**d). The experimental observations from template-assisted NC assemblies are typical signatures of SF, and therefore we infer that SF domains are formed within the template-assisted NC assemblies.

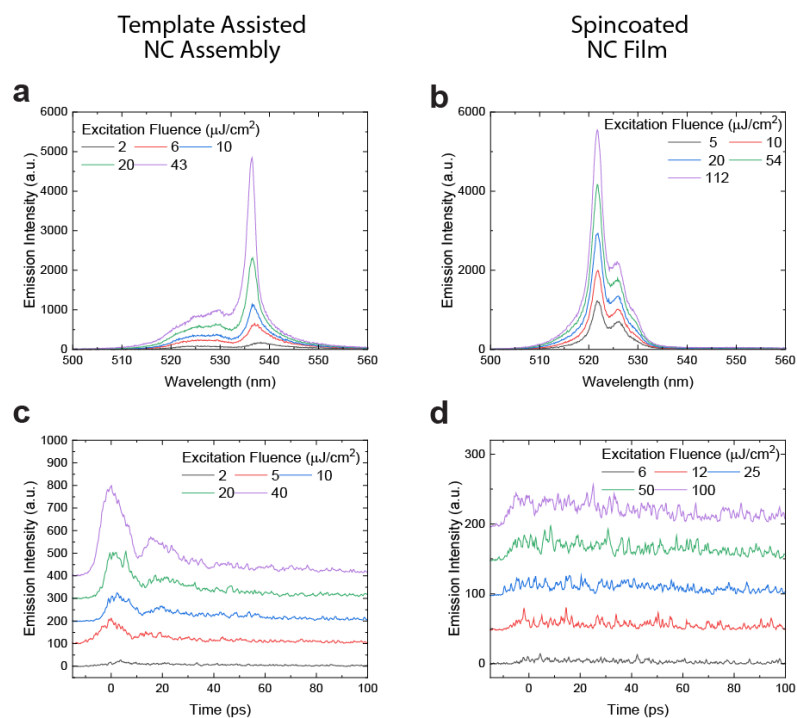

**Figure 6.** Ultrafast spectroscopy under strong femtosecond excitation. (a-b) Time-integrated spectra of a template-assisted NC assembly (a) and a spincoated NC film (b). (c-d) Spectrally-integrated emission time traces for different excitation fluences of a template-assisted NC assembly (c) and a spincoated NC film (d). Each time trace is offset by additional 100 counts (c) or 50 counts (d) in order to visually separate the curves.

**Conclusions**

In conclusion, we developed a methodology to control the size and position of NC assemblies with prefabricated, hollow, transparent template structures. We established a combination of ligands, solvent, and template structures to reliably obtain CsPbBr$_3$ NC assemblies with the template-assisted method. From spectroscopic measurements at cryogenic temperatures, we observed distinct emission dynamics like lifetime shortening, super-linear increase of time-resolved emission peak intensity, and ringing by increasing the excitation fluence, all typical signatures of SF. Our results open the path towards controlled integration of superfluorescent NC assemblies into photonics nanostructures, such as waveguides and microcavities, that will be pivotal for applications.

**Methods**

**Synthesis of CsPbBr$_3$ nanocrystals.** CsPbBr$_3$ NCs have been synthesized according to the recently developed approach with some modifications[24].

***Stock solutions.*** *PbBr$_2$-TOPO stock solution (0.04 M).* PbBr$_2$ stock solution was prepared by dissolving PbBr$_2$ (1 mmol, 376 mg, Sigma-Aldrich) and trioctylphosphine oxide (5 mmol, 2.15 g, TOPO 90%, Strem) in octane (5 mL, Roth) at 100 ºC, followed by dilution with hexane (20 mL) and filtering through a 0.2 µL PTFE filter before use.

*Cs-dopa stock solution (0.02M)* was prepared by mixing 100 mg of Cs$_2$CO$_3$ (Sigma-Aldrich) together with 1 ml of diisooctylphosphinic acid (dopa, Sigma-Aldrich) and octane (2 mL, Roth) at 100 ºC followed by dilution with hexane (27 mL) and filtering through a 0.2 µL PTFE filter before use.

*Didodecyldimethylammonium bromide (DDAB) solution (100 mg/mL, 0.215 M)* was prepared by dissolving 300 mg of DDAB (Sigma-Aldrich) in 3 mL anhydrous toluene (Sigma-Aldrich).

*Oleic acid/oleylamine stock solution* (OAc/OAm) was prepared by mixing 0.632 mL dried oleic acid (Sigma-Aldrich) and 0.66 mL distilled oleylamine (Strem) in 5 mL anhydrous hexane.

*Phosphatidyl-L-serine stock solution* (Ptd-L-Ser, 0.05M) was prepared by mixing 39.6 mg Ptd-L-Ser with 1 mL distilled mesitylene (Acros).

***Synthesis.*** In a 25-ml one-neck flask, 2 mL PbBr$_2$-TOPO stock solution was combined along with 3 mL hexane. Under vigorous stirring, 1 mL of Cs-dopa stock solution was swiftly injected. In 2 min 30 s, a stock solution of ligands (0.15 mL DDAB in toluene or 1.2 mL OAc/OAm-solution together with 35 µL of DDAB in toluene or 0.4 mL Ptd-L-Ser in mesitylene) is added to initiate the ligand exchange on the NC surface. In 2 min 30 s after addition of the ligands, the crude solution was concentrated by evaporating hexane on a rotary evaporator down to less than 1.2 ml of residual solvent. The NCs were precipitated from the concentrated colloid by adding non-solvent.

***Purification***. <u>DDAB-capped</u> NCs were purified using acetone (crude solution:non-solvent 1:1, v/v), followed by solubilization of the obtained NCs in cycloheptane. For <u>OAc/OAm/DDAB-capped NCs</u>, ethyl acetate was used (crude solution:non-solvent, 1:1, v/v), followed by solubilization of the obtained NCs in anhydrous toluene. <u>Ptd-L-Ser-capped</u> NCs were purified using a mixture of ethyl acetate and acetonitrile with a volume ratio of 2:1 (crude solution:non-solvent, 1:3, v/v), followed by solubilization of the obtained NCs in anhydrous toluene.

The concentrations of NCs are about 15 mg/mL (OAc/OAm/DDAB-capped), 10 mg/mL (DDAB-capped ), and 6 mg/mL (Ptd-L-Ser). These solutions were used further for the preparation of the 3D superlattices.

**Absorption and photoluminescence measurements.** UV-VIS absorption spectra were collected using a Jasco V770 Spectrometer operated in transmission mode. A Fluoromax 4 Horiba Jobin Yvon spectrofluorimeter equipped with a PMT detector was used to acquire steady-state PL spectra from solutions. The excitation wavelength was 400 nm, provided by a 150W Xenon lamp dispersed with a monochromator. Measured intensities were corrected to take into account the spectral response of the detector.

**Transmission electron microscopy.** TEM images were collected using a JEOL JEM-2200FS microscope operated at 200 kV.

**Fabrication of 3D hollow template structures.** Silicon-on-insulator (SOI) substrates were treated under oxygen plasma for 5 min to obtain better adhesion of Hydrogen silsesquioxane (HSQ, Dow Corning). HSQ was spin-coated on the substrates. After e-beam exposure and development of HSQ, the lithographically defined pattern of the HSQ was transferred to the Si device layer by inductively coupled plasma reactive ion etching (RIE) with HBr. A 200 nm thick $SiO_2$ layer was deposited on the substrates with atomic layer deposition (ALD) and e-beam evaporated deposition. The openings of the template structures were defined by e-beam lithography with CSAR (Allresist) as resist. After the development of CSAR, the pattern was transferred to the $SiO_2$ layer by RIE with $CHF_3$ and Ar. Lastly, the exposed Si layer was dry-etched with $XeF_2$ gas.

**Template-assisted NC assembly.** 5 – 10 µL of NC solution was drop-casted on a 10 × 10 mm substrate with template structures. The substrate was placed in a petri dish (diameter: 90 mm, height: 20 mm) with 1 mL of toluene. The petri dish was covered by a watch glass so that the solvent evaporated slowly. The evaporation of solvent usually takes 24 hours. For the removal of excess NCs, optics cleaning polymer (Red First Contact Polymer, Photonic Cleaning Technologies) was applied on the substrate surface and peeled off after the polymer solution was dried out.

**STEM cross section imaging**. Template structures with NCs were sliced down to lamella structures with 50 to 80 nm thickness by focused ion beam, using a FEI Helios Nanolab 450S. The lamellas were investigated with a double spherical aberration corrected transmission electron microscope (JEOL ARM200F) operated at 200 kV. Energy dispersive spectroscopy was carried out with liquid-nitrogen free silicon drift detector.

**Optical image analysis.** Optical images of template structures were taken with a digital microscope (Keyence VHX-7000). Quantitative image analysis was done by thresholding the RGB values to discern NC from other structures and counting the unmasked pixels.

**Optical spectroscopy.** For spatial resolved PL map measurements, we used a CW diode laser with the wavelength of 405 nm as the excitation source. The excitation laser was coupled to a single-mode fiber and was focused on the sample with a 100× objective lens (Mitutoyo) to a Gaussian spot diameter with 2.5 μm FWe$^{-2}$M. The sample was mounted on *XYZ* nano-positioning stages. The sample was scanned over 15 × 15 μm area with 1 μm step size.

Time-integrated and time-resolved PL at cryogenic temperature was measured by mounting the samples in a cold-finger flow cryostat which operates down to a temperature of 6 K. We used a fiber-coupled diode laser (PicoQuant) with the wavelength of 405 nm as the excitation source. The laser has a pulse duration of ~50 ps with a repetition rate of 250 kHz. The laser emission was filtered with a short pass filter edge of 430 nm, and was focused on the sample with a 100× objective lens (Mitutoyo) to a Gaussian spot with $1/e^2$ diameter of 4.4 μm. The emitted light from the sample was collected by the same objective and passed through a long-pass filter (Semrock) with 450 nm edge wavelength. The collected light was dispersed by a 300 lines/mm grating in a 0.75 m-long monochromator, and the spectra were recorded by an EMCCD (Princeton Instruments). The PL time traces were recorded with an avalanche photodiode (MPD, time resolution of 50 ps) connected to a time-correlated single photon counting system.

For ultrafast time-resolved measurements with higher time resolution, the samples were mounted in a helium exchange-gas cryostat which operates at a temperature down to 6 K. As an excitation source, we used a frequency-doubled regenerative amplifier running at 400 nm with a repetition rate of 1 kHz, delivering pulses of about 100-200 fs duration. To prevent parasitic light, a short-pass filter (edge wavelength of 492 nm) was used. For both excitation and detection, we used the same focusing lens with 100 mm focal length, resulting in an excitation spot size of about 80 μm in $1/e^2$ diameter. The recorded photoluminescence was spectrally filtered by a long-pass filter (edge wavelength of 460 nm). For the time-resolved measurements, the emission was dispersed by a 150 lines/mm grating in a 0.3-m-long monochromator and detected with a streak camera (Hamamatsu) with a nominal time resolution of 2 ps and measured Gaussian-shaped instrument response function FWHM of 4 ps. The time-integrated photoluminescence spectra were recorded by a 0.5-m spectrograph equipped with a 300 lines/mm grating and a liquid nitrogen-cooled CCD camera.

## ASSOCIATED CONTENT

**Supporting Information**. In the Supporting Information, we present the PL and absorption spectra of NCs, a TEM image of NCs, size distribution of NCs, STEM images and EDS measurement results of a template assisted NC assembly, masked optical microscope images of template assisted NC assemblies for image analysis, detailed statistical analysis of the yield of the template assisted assembling method, and ultrafast PL dynamics of a drying mediated NC assembly before and after applying the cleaning polymer.

Supporting_Information.pdf

## AUTHOR INFORMATION

**Corresponding Author**

Correspondence should be addressed to Maryna I. Bodnarchuk (maryna.bodnarchuk@empa.ch), Maksym V. Kovalenko (mvkovalenko@ethz.ch) and Thilo Stöferle (tof@zurich.ibm.com).

**Funding Sources**

Swiss National Science Foundation, European Commission.

**Notes**

The authors declare no competing interest.

ACKNOWLEDGMENT

   We thank the team of the IBM Binnig and Rohrer Nanotechnology Center for support with the sample fabrication. We thank Armin Knoll, Heiko Wolf, and Heinz Schmid for stimulating discussions. This work was supported by the Swiss National Science Foundation (Grant Number 200021_192308, "Q-Light"), the European Union's Horizon 2020 program through a EIC Pathfinder Open research and innovation action (Grant Agreement No. 899141, "PoLLoC").

**Supporting Information**

# Perovskite nanocrystal self-assemblies in 3D hollow templates

*Etsuki Kobiyama[1], Darius Urbonas[1], Maryna I. Bodnarchuk[2,3], Gabriele Rainò[2,3], Antonis Olziersky[1], Daniele Caimi[1], Marilyne Sousa[1], Rainer F. Mahrt[1], Maksym V. Kovalenko[2,3], Thilo Stöferle[1].*

[1] IBM Research Europe – Zurich, Rüschlikon, Switzerland

[2] Institute of Inorganic Chemistry, Department of Chemistry and Applied Bioscience, ETH Zurich, Zurich, Switzerland

[3] Laboratory of Thin Films and Photovoltaics, Empa — Swiss Federal Laboratories for Materials Science and Technology, Dübendorf, Switzerland

Corresponding Authors: maryna.bodnarchuk@empa.ch, mvkovalenko@ethz.ch, tof@zurich.ibm.com

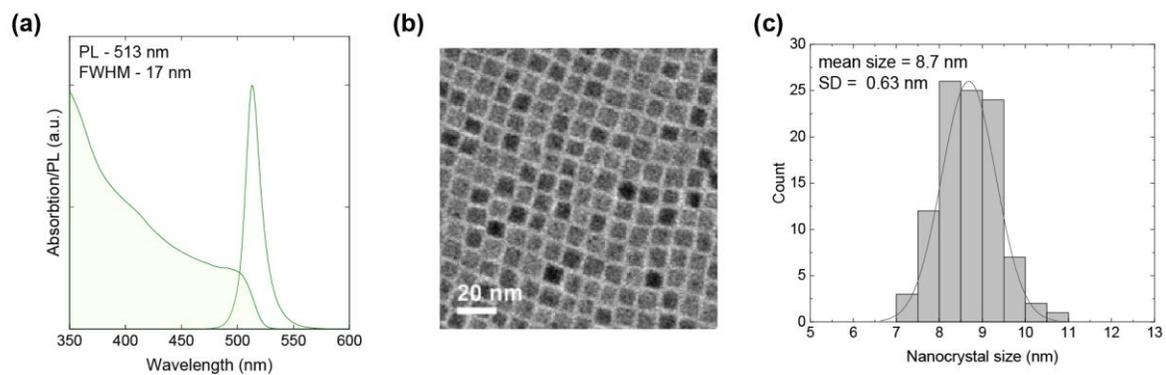

**Figure S1.** (a) Absorption and PL spectra of CsPbBr$_3$ NCs. (b) TEM image of CsPbBr$_3$ NCs. Scale bar: 20 nm. (c) Histogram of edge size of CsPbBr$_3$ NCs.

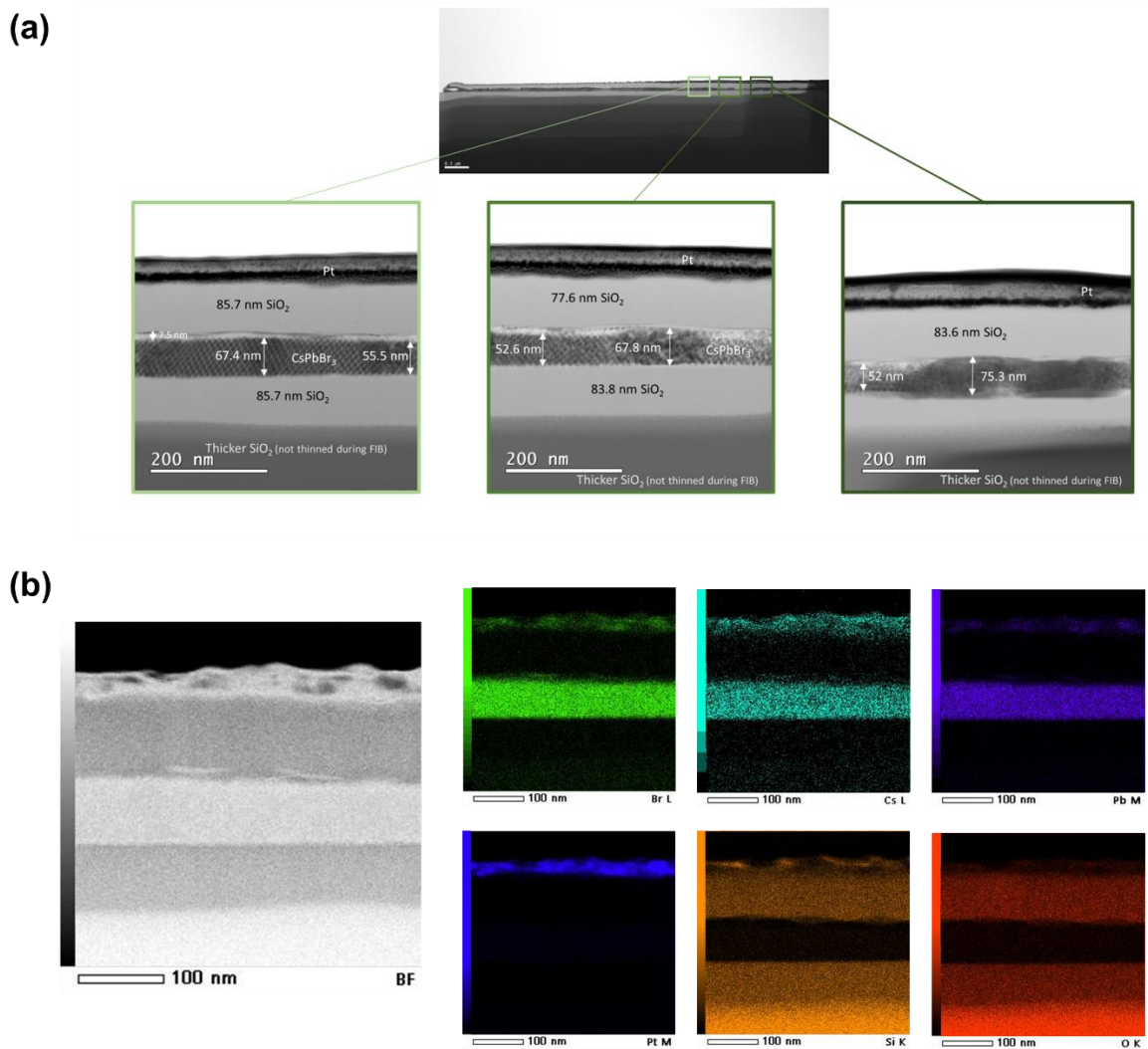

**Figure S2.** (a) Cross section bright field STEM images of different positions of a template assisted NC assembly, revealing grains of various thickness. (b) Results of energy dispersive spectroscopy (EDS). The left panel shows a bright field image of the cross section of a template assisted NC assembly. The right panels display EDS signals of different species: Br (top left, green), Cs (top center, cyan), Pb (top right, violet), Pt (bottom left, blue), Si (bottom center, orange), and O (bottom right, red).

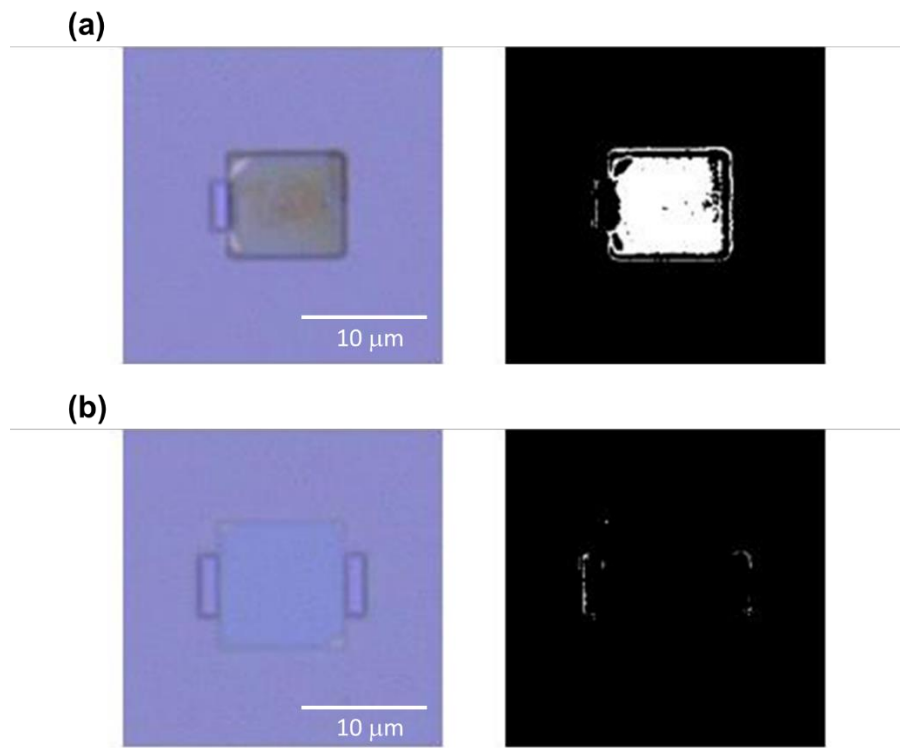

**Figure S3.** (a, b) Typical optical microscope image (left) and the result of binary masking the image (right) of a filled template (a) and an empty template (b).

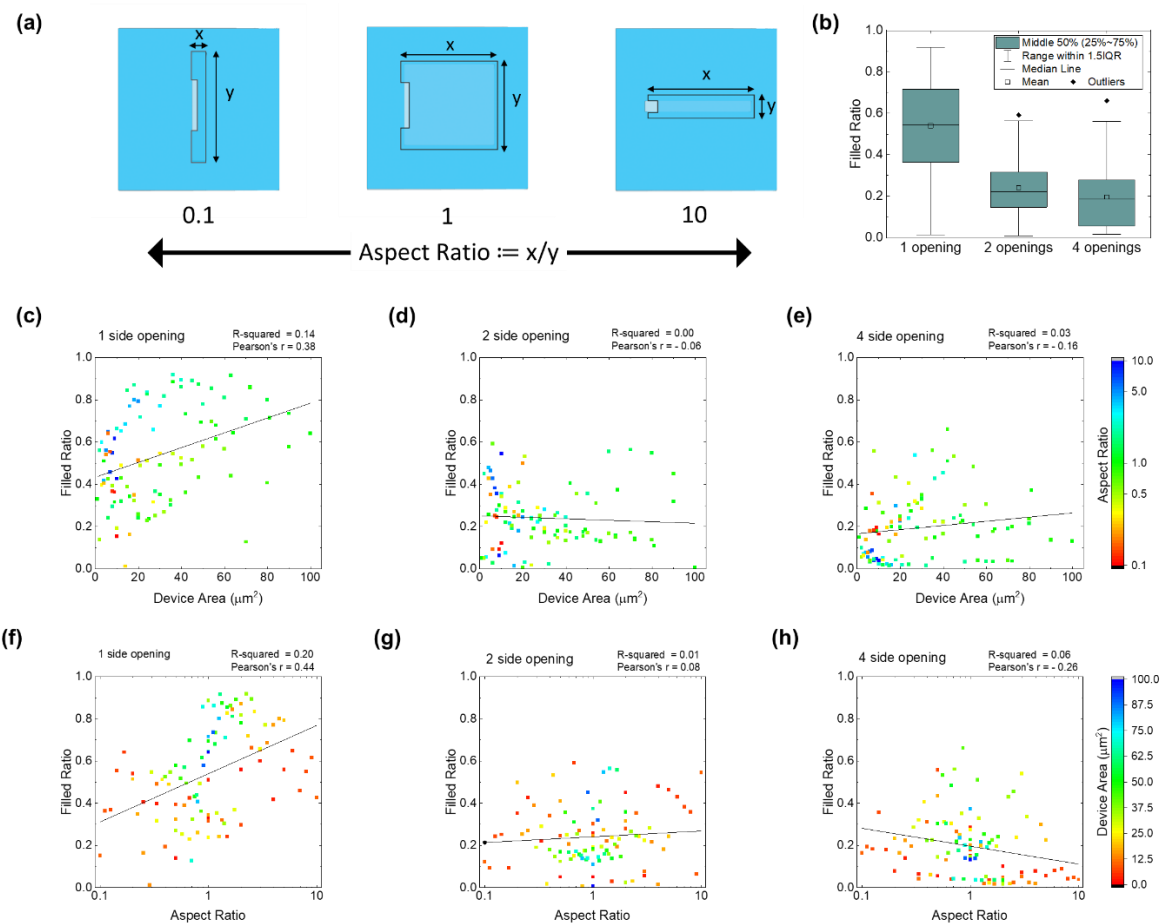

**Figure S4**. (a) Definition of aspect ratios of template structures. (b) Box plots summarizing the distribution of filled ratios of different design template structures. (c-e) Scatter plots of filled ratio against device area for different numbers of openings: (c) on one side, (d) on two sides, and (e) on four sides. (f-h) Scatter plots of filled ratio against aspect ratio of devices for different numbers of openings: (f) on one side, (g) on two sides, and (h) on four sides

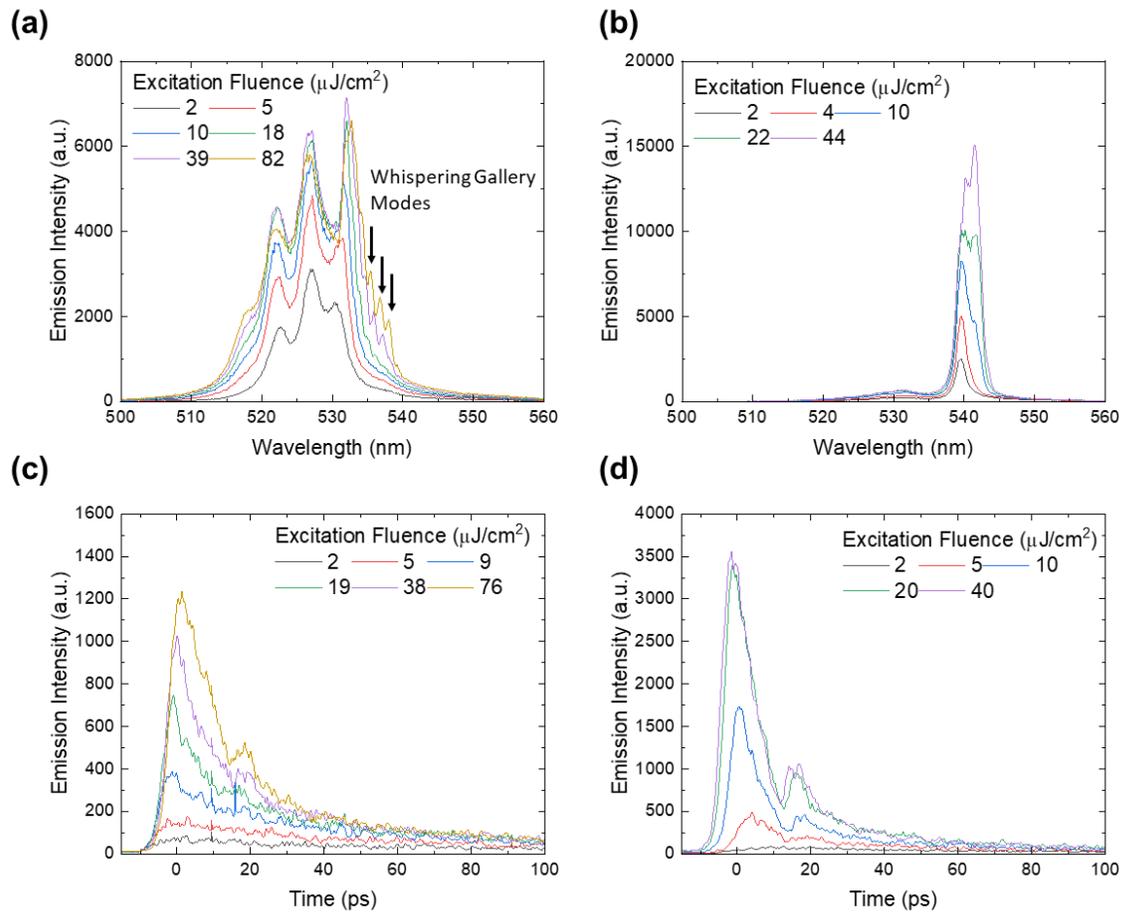

**Figure S5.** Ultrafast spectroscopy under strong femtosecond excitation. (a-b) Time-integrated spectra of a drying mediated NC assembly (a) and a drying mediated NC assembly after applying the cleaning polymer (b). (c-d) Spectrally-integrated emission time traces for different excitation fluences of a drying mediated NC assembly (c) and a drying mediated NC assembly after applying the cleaning polymer (d).